\documentclass[%
preprint,
 amsmath,amssymb,
 aps,superscriptaddress
]{revtex4-2}

\usepackage{graphicx}% Include figure files
\usepackage{bm}% bold math

\newcommand{\fss}{FeSe$_{1-x}$S$_x$ }
\newcommand{\tc}{$T_{\mathrm c}$ }
\newcommand{\musr}{$\mu$SR }

\begin{document}
%\preprint{APS/123-QED}

%\linenumbers

\title{Sulfur-induced magnetism in FeSe$_{1-x}$S$_x$ thin films on LaAlO$_3$ revealed by muon spin rotation/relaxation}

\author{F. Nabeshima}
 \email{cnabeshima@g.ecc.u-tokyo.ac.jp}
\affiliation{Department of Basic Science, the University of Tokyo, Meguro, Tokyo 153-8902, Japan}
\author{Y. Kawai}
\affiliation{Department of Engineering and Applied Sciences, Sophia University, Chiyoda, Tokyo 102-8554, Japan}
\author{N. Shikama}%
\affiliation{Department of Basic Science, the University of Tokyo, Meguro, Tokyo 153-8902, Japan}
\author{Y. Sakishita}%
\affiliation{Department of Basic Science, the University of Tokyo, Meguro, Tokyo 153-8902, Japan}
\author{A. Suter}
\affiliation{Labor f\"{u}r Myonspinspektroskopie, Paul Scherrer Institut, CH-5232 Villigen PSI, Switzerland}
\author{T. Prokscha}
\affiliation{Labor f\"{u}r Myonspinspektroskopie, Paul Scherrer Institut, CH-5232 Villigen PSI, Switzerland}
\author{S. E. Park}
\affiliation{Department of Engineering and Applied Sciences, Sophia University, Chiyoda, Tokyo 102-8554, Japan}
\author{S. Komiya}
\affiliation{Central Research Institute of Electric Power Industry, Yokosuka, Kanagawa 240-0196, Japan}
\author{A. Ichinose}
\affiliation{Central Research Institute of Electric Power Industry, Yokosuka, Kanagawa 240-0196, Japan}
\author{T. Adachi}
\affiliation{Department of Engineering and Applied Sciences, Sophia University, Chiyoda, Tokyo 102-8554, Japan}
\author{A. Maeda}%
\affiliation{Department of Basic Science, the University of Tokyo, Meguro, Tokyo 153-8902, Japan}

\date{\today}% It is always \today, today,

\begin{abstract}
Muon spin rotation/relaxation measurements were performed to investigate magnetic properties of FeSe$_{1-x}$S$_x$ thin films on LaAlO$_3$. 
A drastic decrease of the initial asymmetry was observed together with the peak structure in the temperature dependence of the relaxation rate of muon spins almost at the same temperature where kink anomalies were observed in the temperature dependent resistivity.
With increasing S content, the anomaly temperature increased and the magnetic fluctuations at the lowest temperature were suppressed.
These results show that the S substitution induces magnetism at low temperatures in FeSe$_{1-x}$S$_x$ thin films.
Although the behaviors of the magnetic and nematic phases in FeSe films towards chemical pressure by S substitution are similar to those for bulk FeSe towards hydrostatic pressure, the behavior of \tc is significantly different between these systems. 
Our results demonstrate that the detailed comparative investigation among physical and chemical pressure effects is essentially important to understand the interplay of the magnetism, the nematicity and the superconductivity in iron chalcogenides.
\end{abstract}

\maketitle

\section{\label{s:introduction}Introduction}
Since the discovery of the iron-based superconductors (FeSCs)\cite{Kamihara08}, much research has been devoted to reveal the mechanism of superconductivity in these materials.
The elucidation of the interplay between the nematic order, antiferromagnetic order, and superconductivity in FeSCs has been believed to provide a clue to understand the mechanism of superconductivity in these emergent systems. 
An iron chalcogenide superconductor, FeSe\cite{Wu08}, which is one of the FeSCs with the simplest crystal structure, shows a structural transition from the tetragonal to orthorombic phase at 90 K\cite{PhysRevLett.103.057002}, below which an orbital ordered state was observed\cite{PhysRevB.90.121111,PhysRevLett.113.237001}.
%An iron-chalcogenide superconductor FeSe with the superconducting transition temperature, \tc $\sim 9$ K, exhibits the structural transition from the tetragonal to orthorhombic phase at 90 K, below which an orbital ordered state is observed. %%\cite
%The structural transition is considered to have an electronic origin, and is also called the electronic nematic transition.
Although iron pnictides show antiferromagnetic order near the nematic transition temperature, FeSe exhibits no long-range magnetic order at ambient pressure.
While the superconducting transition temperature, $T_{\mathrm c}$, of bulk FeSe is $\sim$9 K at ambient pressure, it shows a remarkable tunability.
\tc increases up to approximately 40 K by applying hydrostatic pressure\cite{Mizuguchi08,NatMat.8.630}, intercalation\cite{PhysRevB.82.180520,SciRep.2.426,JPSJ.82.123705,Shi_2018} and electron doping by ionic gating\cite{PRL.116.077002}.
A recent study reported the observation of zero resistivity up to 46 K in an FeSe and ionic liquid system\cite{ShikamaAPEX2020}.
Further enhancement of \tc up to 65-100 K was reported in FeSe monolayer films on SrTiO$_3$\cite{CPL.29.037402,NatMater.14.285}.

Applying hydrostatic pressure to FeSe induces an antiferromagnetic order at low temperatures after the nematic order was almost suppressed (at approximately 2 GPa)\cite{PhysRevB.85.064517,JPSJ.84.063701,NatCom.7.12146,NCom.7.12728}. %他の論文も
The magnetic phase shows a dome shape in the pressure-temperature phase diagram peaked at approximately 4 GPa.
$T_{\mathrm c}$ significantly increases with increasing applied pressure, and reaches the maximum of approximately 40 K at 6 GPa.
In the case of chemical pressure application, isovalent S substitution for Se also suppress the nematic order\cite{PNAS.113.8139,PRB.96.121103}, but antiferromagnetic order has not been reported for FeSe$_{1-x}$S$_x$ at ambient pressure\cite{NatCom.8.1143,PhysRevB.96.024511,PhysRevLett.123.147001}.  %引用他の論文も
%Matsuura $et\ al$. attributed the difference in the phase diagrams of FeSe between physical and chemical pressure to the difference in the chalcogen height from the two-dimensional Fe plane\cite{NatCom.8.1143}.
%Recently, the growth of superconducting thin films of FeSe$_{1-x}$S$_x$ was reported\cite{Nabe18.FeSeS}.
However, thin film samples of FeSe$_{1-x}$S$_x$ shows a kink anomaly in the temperature dependence of resistivity, $\rho$, for $x>0.2$, where the nematic order was completely suppressed\cite{Nabe18.FeSeS}.
The kink anomaly was similar to those observed in FeSe under hydrostatic pressure, which was attributed to the antiferromagnetic transition\cite{JPSJ.84.063701}.
Very recently, hydrothermal synthesis of bulk FeSe$_{1-x}$S$_x$ single crystals with $x$ covering the full range ($0 \leq x \leq 1$) was reported for the first time\cite{arXiv:2010.05191}.
The FeSe$_{1-x}$S$_x$ bulk crystals with high S content ($0.31 \leq x \leq 0.71$) shows similar resistivity upturns at low temperatures.
%The resistivity upturn is most prominent for $x=0.45$, with the upturn temperature showing a dome-like $x$ dependence.
These resistivity anomalies observed in heavily-S-substituted samples for both films and bulk suggest the occurrence of the antiferromagnetic order in FeSe$_{1-x}$S$_x$.

In this article, we report muon spin rotation/relaxation ($\mu$SR) measurements for FeSe$_{1-x}$S$_x$ thin films.
We observed a significant decrease of the initial asymmetry at low temperatures and also observed a peak structure in the temperature dependence of the relaxation rate of muon spins at the same temperature, which coincides with the temperature of the kink anomalies in the $\rho$-$T$ curve.
%is also the same as the temperature of the kink anomalies in the $\rho$-$T$ curve.
These observations suggest that FeSe$_{1-x}$S$_x$ thin films exhibit a magnetic transition at low temperatures. %, probably due to the lattice strain existing only in film samples.  
Our findings demonstrate that the detailed comparative investigation among physical and chemical pressure effects for bulk and film samples is essentially important to understand the interplay of the magnetism, the nematicity and the superconductivity in iron chalcogenides.
%Our findings demonstrate that FeSe$_{1-x}$S$_x$ films will be      to understand interplay of the magnetism, the nematicity and the superconductivity in iron chalcogenides.%

\section{\label{s:experimental}Experimental}

%\subsection{\label{s2:sample}Sample preparation}
Single crystalline thin films of FeSe$_{1-x}$S$_x$ with $x=$0.3 and 0.4 were grown on LaAlO$_3$ (LAO) substrates by a pulsed laser deposition method using a KrF laser\cite{Imai09,Imai10}.
Details of the film growth and the estimation of the composition was published elsewhere\cite{Nabe18.FeSeS}.
The maximum size of the samples that we are able to grow at the same time was 15 mm $\times$ 15 mm.
We grew films with the same composition twice to obtain samples with size of 20 mm $\times$ 20 mm necessary for the \musr measurements.
The crystal structures and the orientations of the films were characterized by four-circle X-ray diffraction (XRD) with Cu K$\alpha$ radiation at room temperature.
The electrical resistivity was measured by a standard four-probe method using a physical property measurement system from 2 to 300 K.
The thicknesses of the grown films were estimated from the electrical resistance at room temperature assuming the resistivity at room temperature ($\rho =$ 0.5 m$\Omega$cm for both compositions), which typically has variations of approximately 20\% at most. % 20%程度のエラーあり

%\subsection{\label{s2:musr}Muon spin rotation (\musr) measurement}

The \musr measurements for the thin films were performed on the $\mu$E4 beamline at the Paul Scherrer Institut, Switzerland using the low-energy muon. %at the Swiss Muon Source $\mu$ at the Paul Scherrer Institut, Switzerland using the low-energy \musr facility LEM.
The thin film samples, covering a 20$\times$20 mm$^2$ area, were glued using Ag-paste onto a Ni-coated aluminium plate, which was then mounted onto a cryostat.
The spin relaxation of muons that stop in the Ni-coated plate is very fast and their contributions to the measured asymmetry are negligible for $t >$ 0.05 $\mu$sec\cite{SAADAOUI2012164}.
The implanted beam energy, $E_{\mathrm {imp}}$, was chosen as 3 keV and 4 keV for $x=0.3$ and $x=0.4$, respectively, so that the muons stopped in the center of the films for both compositions.
%温度
%Transverse field longitudinal field zero field 
The spin of the incident muons is parallel to the film plane for the zero-field (ZF) and weak-transverse-field (wTF) experiments and perpendicular to the film plane for the longitudinal-field (LF) experiments.
The magnetic field was applied perpendicular to the film plane for both the wTF and LF experiments.
Details concerning the LEM setup can be found elsewhere\cite{MORENZONI2000653,PROKSCHA2008317}. %
All the \musr data has been analyzed with the help of musrfit\cite{SUTER201269}.

\section{\label{s:results}Results}

\begin{figure*}[htb]
\includegraphics[width=0.95\linewidth]{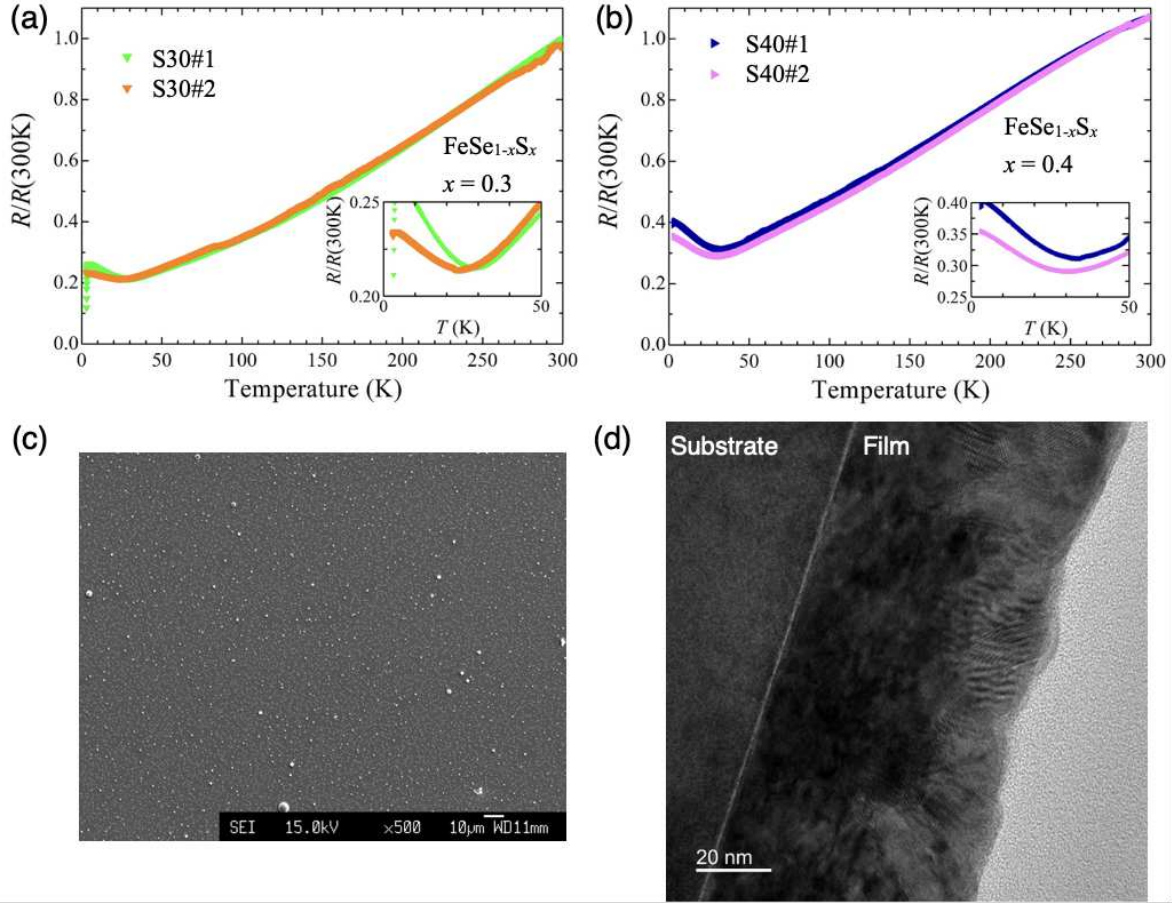}
\caption{(a),(b) Temperature dependence of the normalized dc electrical resistivity of the FeSe$_{1-x}$S$_x$  films on LAO with (a) $x=0.3$ (S30$\#$1 and S30$\#$2) and (b) $x=0.4$ (S40$\#$1 and S40$\#$2). (c) SEM image for a film with $x=0.3$. (d) Cross-sectional TEM image for a film with $x=0.4$. This image was taken at a place where there was no precipitates at the surface.}
\label{G-RT}
\end{figure*}

The specifications of the measured films are summarised in table \ref{tab:FilmSpec}.
Because we grew the films in two parts for each composition, both samples were characterized, which are named as \#1 and \#2, respectively.
In the \musr experiments the two samples were mounted on the sample holder simultaneously and measured.
All the films have shorter $c$-axis lengths compared with bulk samples with the same composition ($c \sim 5.39$~\AA ~for $x=0.3$ and $c \sim 5.34$~\AA ~for $x=0.4$)\cite{PhysRevB.98.020507}, suggesting tensile strain in the films.
Figure \ref{G-RT}(a) and (b) shows the temperature dependence of the electrical resistance, $R$, of the films. % which were grown at the same time as the films used for the \musr measurements.
The two samples with the same composition showed the almost identical temperature dependence of the resistance for both $x=$0.3 and 0.4.
Some samples showed the signature of the superconducting transition at the lowest temperature, but did not show zero resistance above 2 K.
The \tc values of the films are lower than bulk samples\cite{arXiv:2010.05191}, which might be attributed to the tensile strain in the films\cite{Nabe13,yi15pnas,Nabe18.FeSeStrain}. 
The kink anomaly was observed in the $R$-$T$ curves for both $x=$0.3 and 0.4. 
The temperature of the kink anomaly, $T_{\mathrm {kink}}$, increased with increasing S content, as was reported in the previous work\cite{Nabe18.FeSeS}.
Note that no signature of the nematic transition was observed in the temperature dependence of resistance for these samples with high S contents.  Even if there is any nematic-ordered domains in the samples, they do not affect the \musr time spectra unless strong magnetic field is applied parallel to the $ab$ plane\cite{NatMater.14.210}.

% 0.3  5.39 \AA   0.4 5.34

Figure \ref{G-RT}(c) shows a scanning electron microscope (SEM) image for a film with $x=0.3$.
A lot of tiny precipitates smaller than 1 $\mu$m were observed.
These precipitates were too small for the electron beam to be focused on one of them, but the composition analysis with different beam sizes revealed that the precipitates had a larger proportion of Fe than that of the underlying film. 
Figure \ref{G-RT}(d) shows a cross-sectional transmission electron microscope (TEM) image for a film with $x=0.4$.
This image was taken at a place where there were no precipitates at the surface.  
Although $c$-axis orientation was observed near the interface between the film and the substrate, there are domains with different orientations near the surface, which may be consistent with the fact that the XRD peak intensity became weaker for films with more S content\cite{Nabe18.FeSeS}. 
Tilted striped patterns of the contrast observed near the surface of the films are Moire fringes due to domains with different orientations.
We confirmed that the composition is homogeneous in the whole films including these domains. 
%組成は変わらない．入射方向．ごく表面近傍には酸化層がある膜厚xx見えづらいが

\begin{figure*}[htb]
\includegraphics[width=0.95\linewidth]{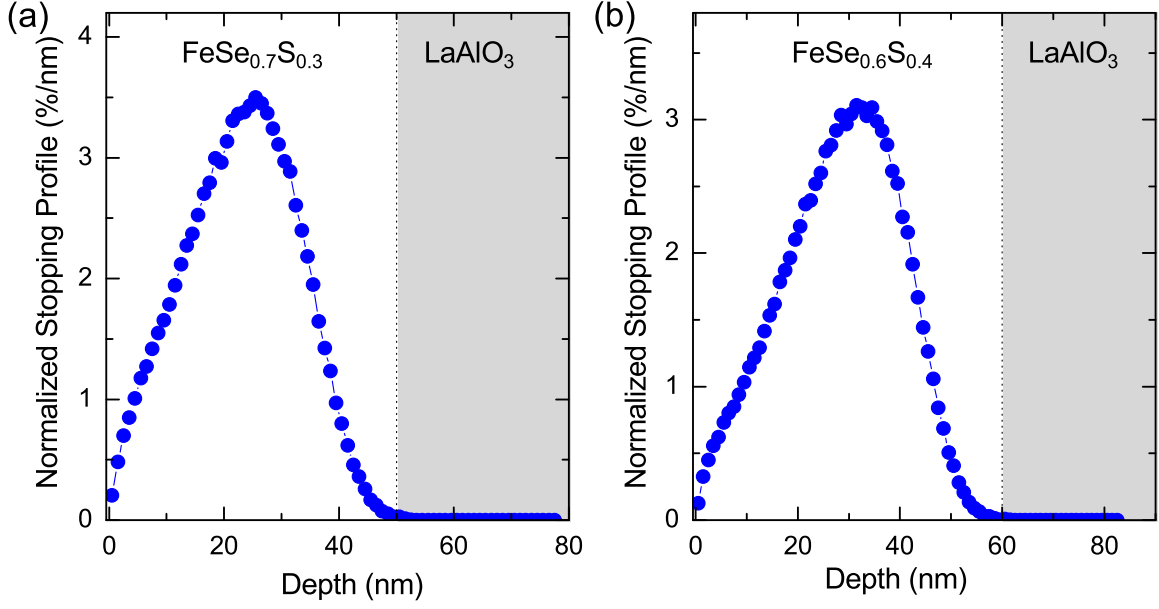}
\caption{Muon stopping profiles from Monte-Carlo calculations for (a) 50-nm-thick FeSe$_{0.7}$S$_{0.3}$ film with $E_{\mathrm {imp}}$ = 3 keV and (b) 60-nm-thick FeSe$_{0.6}$S$_{0.4}$ film with $E_{\mathrm {imp}}$ = 4 keV.}
\label{G-SP}
\end{figure*}

%\subsection{\label{s2:ZF}Stopping Profile}
Figure \ref{G-SP} shows the muon stopping profiles
from Monte-Carlo calculations for the thinner films for each composition (the 50-nm-thick film for $x$ = 0.3 with $E_{\mathrm {imp}}$ = 3 keV and the 60-nm-thick film for $x$ = 0.4 with $E_{\mathrm {imp}}$ = 4 keV).
The modes of the distribution are almost at the center of both films and implanted muons are distributed throughout the samples.  
These results suggest that most of muons stopped deeper below the surface layer where the domains were observed.
The calculation also suggests that there were very few muons that stop inside the substrate.
Even if some muons stop inside the substrate, the LaAlO$_3$ substrate is paramagnetic and the relaxation of the muon spins are very weak.  
Thus, we can safely exclude the possibility that some fraction of the signal arises from substrates.

\begin{figure*}[htb]
\includegraphics[width=0.95\linewidth]{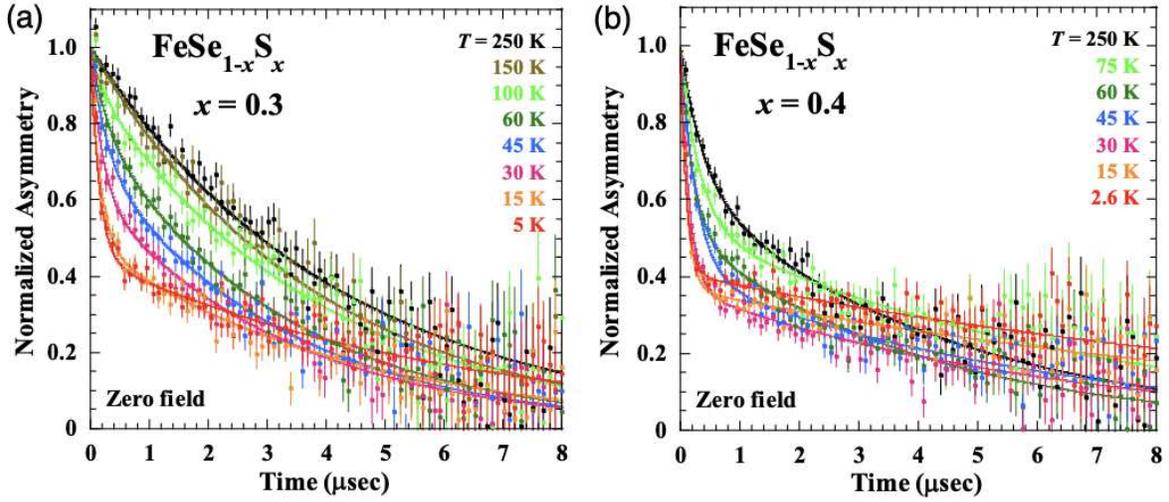}
\caption{Zero-field (ZF) muon time spectra of the FeSe$_{1-x}$S$_x$ thin films with (a) $x=0.3$, and (b) $x=0.4$. Solid lines are the fitted curves obtained by a two-component analysis.}
\label{G-ZF}
\end{figure*}

%\subsection{\label{s2:ZF}ZF spectrum}
Figure \ref{G-ZF} shows the ZF $\mu$SR time spectra of the FeSe$_{1-x}$S$_x$ thin films.
The ZF spectrum at 250 K for $x=0.3$ shows an exponential-type relaxation.
This suggests the development of magnetic correlations near room temperature for $x=0.3$.
As the temperature decreases, the initial relaxation becomes faster.
The asymmetry in the long time region for $x=0.3$ increased with decreasing temperature below 15 K.
This indicates that a nearly static magnetic order was formed at low temperatures.
However, coherent muon precession was not observed even at the lowest temperature, suggesting inhomogeneity in the internal magnetic field.
The slow relaxation in the long time region at the lowest temperature suggests the formation of the static magnetic order with some remaining spin fluctuations.

The ZF spectrum for $x=0.4$ at 250 K also exhibits an exponential-like relaxation and the relaxation is much faster than $x=0.3$, suggesting the magnetic correlation at high temperatures is much stronger than that for $x=0.3$.
The initial decrease of the asymmetry became fast with decreasing temperature, and the asymmetry in the long time region was larger at 2.6 K than at 30 K. 
The relaxation of the ZF spectrum at the lowest temperature in the long time region was slower than that for $x=0.3$, suggesting weaker spin fluctuations in $x=0.4$.

%The ZF spectrum at 250 K for $x=0.4$ shows an exponential-type relaxation.
%This suggests the development of magnetic correlations even near room temperature for $x=0.4$.
%As the temperature decreases, the initial relaxation becomes faster.
%The asymmetry in the long time region for $x=0.4$ increased with decreasing temperature below 30 K.
%This indicates that a nearly static magnetic order was formed at low temperatures.
%However, coherent muon precession was not observed even at the lowest temperature, suggesting inhomogeneity in the internal magnetic field.
%The slow relaxation in the long time region at the lowest temperature suggests the formation of the static magnetic order with some remaining spin fluctuations.

%The ZF spectrum for $x=0.3$ at 250 K also exhibits an exponential-like relaxation and the relaxation is much slower than $x=0.4$, suggesting the magnetic correlation at high temperatures is much weaker than that for $x=0.4$.
%The initial decrease of the asymmetry became fast with decreasing temperature, and the asymmetry in the long time region was larger at 5 K than at 15 K. 
%The relaxation of the ZF spectrum at the lowest temperature in the long time region was faster than that for $x=0.4$, suggesting larger spin fluctuations in $x=0.3$.

%\subsection{\label{s2:2comp}Two-component analysis for TF spectrum}
To reveal how the magnetic order develops, we performed wTF \musr at 50 G for both samples and applied a two-component analysis.
We assumed that the wTF spectra $A(t)$ are described as
\begin{equation}
\label{eq:TF}
A(t)=(A_0 \exp{(-\lambda _0 t)}+A_1 \exp{(-\lambda _1 t)})\times \cos{(\omega t + \phi)},
\end{equation}
where the first and the second terms represent the slow and fast relaxation components, respectively. % ($\lambda_0 < \lambda_1$). 
$A_0$ and $A_1$ are the initial asymmetries, and $\lambda_0$ and $\lambda_1$ are the relaxation rates of muon spins of the slow and fast components, respectively.
$\omega$ and $\phi$ are the frequency and phase of the muon-spin precession, respectively.
The second term represents a $missing$ asymmetry, causing the reduction of $A_0$.

\begin{figure*}[htb]
\includegraphics[width=0.95\linewidth]{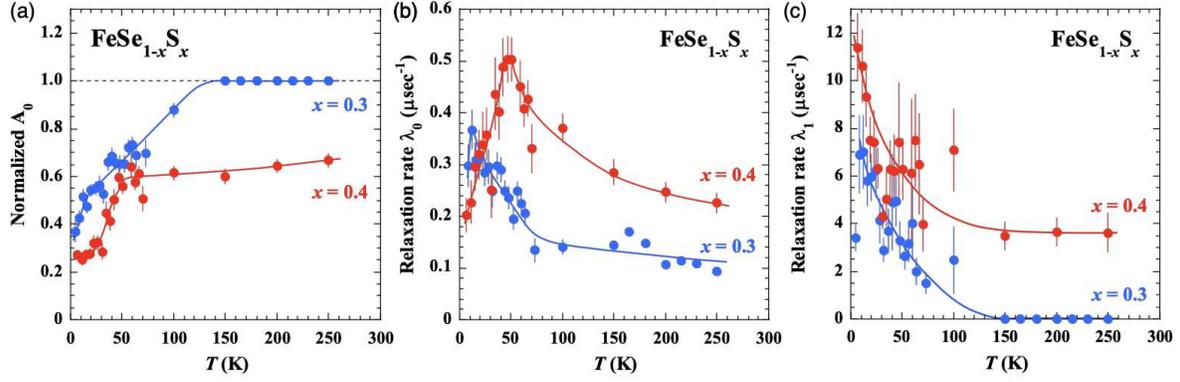}
\caption{Temperature dependence of (a) the initial asymmetry, $A_0$, (b) the relaxation rate of muon spins of the slow component, $\lambda_0$, and (c) the relaxation rate for the fast component, $\lambda_1$, in the equation (\ref{eq:TF}) obtained by the wTF \musr at 50 G for FeSe$_{1-x}$S$_x$ films with $x=$0.3 and 0.4. Solid curves are guides for the eye.}
\label{G-TF}
\end{figure*}

Figure \ref{G-TF}(a) and (b) show the temperature dependence of $A_0$ and $\lambda_0$ of FeSe$_{1-x}$S$_x$ thin films.
The film with $x=0.3$ has no fast relaxation component above 150 K. %\footnote{As a matter of form, the second term $A_1 \exp{(-\lambda _1 t)}$ on the right side of eq. (\ref{eq:TF}) above 150 K can be finite. However, $\lambda_1$ for $x=0.3$ above 150 K is 0.1-0.2 $\mu$sec at most, comparable to $\lambda_0$ and much smaller than that of the slow component for $x=0.4$. Thus, we assumed that there was no fast component above 150 K.}.
%($\lambda_1$ for $x=0.4$ at 250 K is 4$\mu$sec. $\lambda_1$ for $x=0.3$ at 60-100 K is 2$\mu$sec.)
$A_0$ of $x=0.3$ started to decrease below 150 K, and dropped steeply below 20 K.
$\lambda_0$ started to increase steeply below 70 K and took a maximum around 20 K.
%There is no visible feature in $\lambda_0$ at 150 K.
These results suggest that the film with $x=0.3$ shows a magnetic transition around 20 K.

For $x=0.4$, the $A_0$ values at 250 K were already suppressed, which corresponds to the fast exponential relaxation in the ZF data at high temperatures.
This means that stronger magnetic correlations develop already at 250 K compared with the $x=0.3$ sample.
%the magnetic correlation developed at 250 K in some part of the sample.
%, the volume fraction of which was approximately 40\%.
%The $A_0$ values for $x=0.3$ at 40-60 K is close to that for $x=0.4$ above 50 K.
%This may suggest that the development of the magnetic correlation observed for $x=0.3$ below 150 K, corresponding to the gradual decrease of $A_0$, has the same origin as that for $x=0.4$ above 50 K. 
%The temperature dependence of $A_0$ may be interpreted as follows: regions with strong magnetic correlation, which has already developed at 250 K, expand with decreasing temperature and spread throughout the sample at 50 K when they start to interact with each other. 
As the temperature decreased, $A_0$ for $x=0.4$ decreased gradually above 50 K and rapidly dropped below 50 K, and was temperature independent below 20 K. 
The relaxation rate, $\lambda_0$, took a maximum value at 40-50 K, which coincides with the temperature where $A_0$ started to decrease rapidly.
These results indicate that the $x=0.4$ film exhibited a magnetic transition at approximately 50 K, which is higher than that for $x=0.3$.
The facts that no coherent muon precession was observed in the ZF measurements and that $\lambda_0$ remains large even at the lowest temperature suggest the magnetic state at low temperatures is of short-range order for both $x$ = 0.3 and 0.4.

Figure \ref{G-TF}(c) show the temperature dependence of the relaxation rate of the fast component, $\lambda_1$.
$\lambda_1$ for $x$ = 0.3 is zero at room temperature because there is no fast component at high temperatures, while that for $x$ = 0.4 is finite at room temperature.  
For both compositions, $\lambda_1$ monotonically increases with decreasing temperature. 
The increase of $\lambda_1$ is related to the development of the static magnetic order. 
With decreasing temperature, the spin fluctuations in the static-order regions become slow and the muon experiences stronger magnetic moment, which results in enhancement of $\lambda_1$.

In the case of a three-dimensionally isotropic distribution of the internal field, such as in polycrystalline samples, $A_0$ should be 1/3.
On the other hand, in the case of single-crystalline samples, $0<A_0<1$ depending on the angle between directions of the incident muon spins and the internal field.
Therefore, the lowest estimate of the magnetic volume fraction in the films is 65\% and 75\% for $x=$ 0.3 and 0.4, respectively.
Note that there are domains with different orientations near the surface of the films.
Thus, $A_0$ will be larger than 0, and accordingly the magnetic volume fraction will be larger than the above values.
Considering the fact that $A_0$ was temperature independent below 20 K for $x=0.4$, the volume fraction might be almost 100\% for $x=0.4$.

%\subsection{\label{s2:LF}Evaluation of internal field}

\begin{figure*}[htb]
\includegraphics[width=0.75\linewidth]{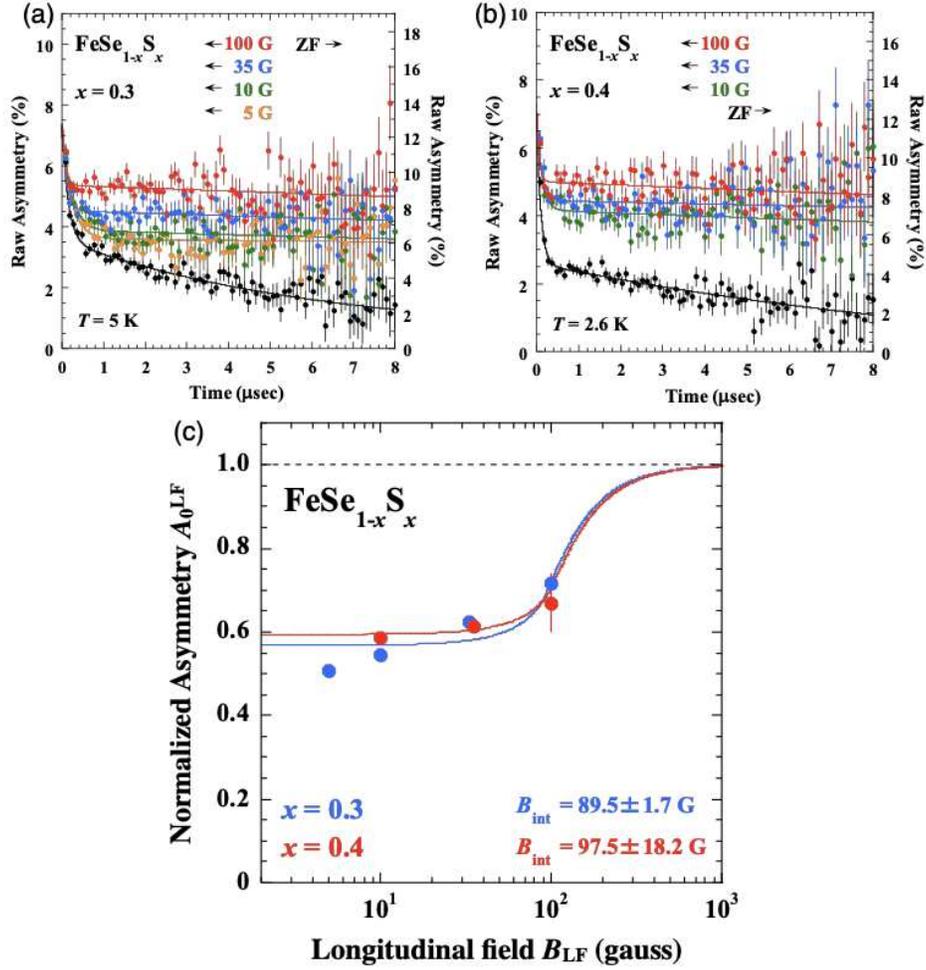}
\caption{(a),(b) LF muon time spectra for (a) $x=0.3$ at 5 K and (b) $x=0.4$ at 2.6 K. Solid lines are the fitted curves with a two-component analysis. (c) The normalized initial asymmetry as a function of applied longitudinal field. Solid lines are the fitting results by the theoretical curves for the case of homogeneous internal magnetic field in samples.}
\label{G-LF}
\end{figure*}

To obtain further information on the low temperature magnetic phase, LF \musr spectra were measured up to $B = 100$ G.
Figure \ref{G-LF}(a) and (b) shows the LF spectra for $x=$ 0.3 and 0.4 at the lowest temperature.
The LF spectra for $x=$ 0.3 showed a parallel upward shift with increasing applied magnetic field.
However, the field of 100 G did not fully recover the initial asymmetry.
If the ZF muon spin relaxation is caused only by nuclear spins, applied field of tens of gauss should completely recover the initial asymmetry.
Thus, the observed LF spectra suggest that there is a static magnetism at the lowest temperature. % in some regions of the sample. 
%The LF spectra for 33 G and 100 G show the gradual decreases of the asymmetry with time.
%This is due to spin fluctuations and suggests the coexistence of the static ordered regions and fluctuating regions.
%At the lowest temperature, the slow relaxation in the long time region vanished and the LF spectra showed parallel upward shift (Fig. \ref{G-LF}(b)), suggesting the increase in the volume of the static ordered region.
%As for $x=0.3$, the recovery of the initial asymmetry by the applied field at 5 K was smaller than that at 250 K.
%This suggests that the internal magnetic field is larger at 5 K.
%Note that a slight relaxation of the asymmetry was observed for $x=0.3$ at 100 G even at the lowest temperature.
%This suggests stronger spin fluctuations in the film with $x=0.3$.
Qualitatively the same behavior was also observed for $x=0.4$.

Figure \ref{G-LF}(c) shows the longitudinal field dependence of the initial asymmetry, $A_0^{\mathrm {LF}}$.
The low-field limit of $A_0^{\mathrm {LF}}$ is not zero, suggesting the internal magnetic field is not parallel to the film plane.
$A_0^{\mathrm {LF}}$'s for both $x=$ 0.3 and 0.4 increased with increasing magnetic field.
In the case that the magnitude of the internal field is homogeneous and the direction of the field is uniformly distributed in space, such as in a polycrystalline sample, the magnetic field dependence of $A_0^{\mathrm {LF}}$ is written as\cite{Pratt_2007}  %homogeneous in magnitude and have random orientation in space 
\begin{equation}
\label{eq:LF}
A_0^{\mathrm {LF}}=\frac{3}{2}(1-A_0^{\mathrm {ZF}}) \biggl[ \frac{3}{4}-\frac{1}{4k^2} +\frac{(k^2-1)^2}{8k^3}\ln{\left|\frac{k+1}{k-1}\right|} \biggr] +\frac{3}{2}(A_0^{\mathrm {ZF}}-\frac{1}{3}),
\end{equation}
where $k$ is the ratio of the applied longitudinal field, $B_{\mathrm {LF}}$, to the internal field, $B_{\mathrm {int}}$ ($k=B_{\mathrm {LF}}/B_{\mathrm {int}}$), and $A_0^{\mathrm {ZF}}$ is the initial asymmetry under the zero magnetic field.
Because we did not evaluated the normalization factor for the LF experiments, which used the different spin polarization of the muon beam and different sets of positron counters from those in the ZF experiments, we fitted the data using this equation with $A_{\mathrm {ZF}}$ as a free parameter.
The solid lines in Fig. \ref{G-LF}(c) are the best fits of the data for both $x=$ 0.3 and 0.4, and the obtained fitting parameter, $B_{\mathrm {int}}$, is $89.5 \pm 1.7$ G and $97.5 \pm 18.2$ G for $x=$ 0.3 and 0.4, respectively. 
Deviations from the fitting curves for both $x=$ 0.3 and 0.4 suggest inhomogeneous spatial distributions of the internal magnetic field in the samples, consistent with no coherent muon precession in the \musr time spectra.
Considering the inhomogeneity of $B_{\mathrm {int}}$, the small difference in the fitted results of $B_{\mathrm {int}}$ between $x$ = 0.3 and 0.4 has no significant difference. 
For more accurate evaluation of $B_{\mathrm {int}}$, measurements with higher magnetic fields are needed.

\section{\label{s:discussion}Discussion}

The \musr experiments revealed that the kink anomalies in the $R$-$T$ curves for FeSe$_{1-x}$S$_x$ thin films originate from a magnetic transition. 
The magnetic transition temperature increases and the magnetic order becomes more stable by the S substitution. 
Our results also suggested the inhomogeneity of the internal field, or the short-range nature of the ordered state.
One may concern that the observed magnetism originates from the precipitates on the surface of the films.
As was described before, the precipitates on the films contain more Fe than the underlying films, and thus, are very likely to be magnetic.
However, the magnetic volume fraction in the FeSe$_{1-x}$S$_x$ films is large and almost 100\% for both $x=$ 0.3 and 0.4, which cannot be attributed to the small magnetic precipitates.
A \musr study on FeS reported that there is a small non-superconducting magnetic phase within a sample\cite{PhysRevB.94.134509}, which may be related to a $\sqrt{5}\times\sqrt{5}$ reconstructed insulating phase observed in a scanning tunneling microscope measurement\cite{ChengweiWang47401}.  
The impurity magnetic phase observed in FeS has very small volume fraction of 15\% and the peak of the relaxation rate is 5 K, which is much lower than the magnetic transition temperatures observed in our measurements.  
These results suggest that the origin of the magnetic order in the Fe(Se,S) films is different from that of the magnetic impurity phase observed in bulk FeS.

One may also concern that small domains with different orientations near the surface of the films are the cause of the spatial inhomogeneity of the internal magnetic field.
However, we should note that no coherent muon precession, which also suggests the inhomogeneity of the internal field, is not explained by the effect of these domains alone, because the muons penetrated deep in the film.
Thus, there is the inhomogeneity of the internal field, at least, in the single crystalline layer underneath.
The origin of the short range nature of the magnetism remains an open question.

Our results may suggest that the resistivity anomalies observed for bulk FeSe$_{1-x}$S$_x$\cite{arXiv:2010.05191} also originate from the magnetic transition.
Interestingly, the composition and the temperature where the resistivity anomaly is observed are different between bulk and film samples.
While the resistivity anomalies of bulk samples are observed for $x>0.3$, those of FeSe$_{1-x}$S$_x$ films are observed for $x > 0.2$, and the anomaly temperatures of films are higher than those of bulk. 
At present, the origin of the difference is not clear, but this may be due to tensile strain in the films.
Studies on strain dependence are now under way. 

%The other possibility is the effect of strain existing in film samples.
%As was described above, while bulk samples of FeSe$_{1-x}$S$_x$ show no magnetism under ambient pressure, bulk FeSe exhibits antiferromagnetism under hydrostatic pressure.
%Matsuura $et\ al$.\cite{NatCom.8.1143} compared their results in bulk FeSe$_{1-x}$S$_x$ with a report on the hydrostatic pressure effects in FeSe by Millican $et\ al$.\cite{MILLICAN2009707}, and attributed the difference in the phase diagrams of FeSe between physical and chemical pressure to the difference in the chalcogen height from the two-dimensional Fe plane.  
%This may suggest that the chalcogen heights in the films are different from those of bulk FeSe$_{1-x}$S$_x$, and are close to the values for FeSe under pressure.
%We should note, however, that some groups reported a different behavior of the chalcogen height in FeSe under hydrostatic pressure other than the results of Millican $et\ al$.
%Although Millican $et\ al$. reported that the chalcogen height was almost pressure independent% while $a$- and $c$-axis length decreased with increasing pressure
%, other groups reported that the chalcogen height also decreased with increasing pressure\cite{Marga09,PhysRevB.81.205119}, similar to the behavior for bulk FeSe$_{1-x}$S$_x$.
%This calls for further detailed studies on the structural analysis for both film and bulk samples to clarify the relation between lattice parameters and the magnetism in FeSe.

%To confirm this, further study on the structure-refinement analysis for the thin film samples is needed. 

From the perspective of the impact of the magnetism on the superconductivity, the $T_{\mathrm c}$ values for the \fss films show gradual and monotonic decrease with increasing the S content even when the magnetic order appears\cite{Nabe18.FeSeS}, which is very similar to that for bulk \fss samples for $x<0.3$, which show no magnetism under ambient pressure.
This may suggest that the magnetism observed in the films has no impact on superconductivity. 
However, comparing with bulk FeSe$_{1-x}$S$_x$ samples with the same composition, the $T_{\mathrm c}$ values of film samples are lower than that of bulk samples by more than 1 K for $x=0.3$\cite{Nabe18.FeSeS,PhysRevB.98.020507}.
In addition, \tc of bulk samples becomes smallest when the resistivity-upturn temperature becomes highest.
These suggest that the magnetism may suppress the superconductivity. 
Because the strain in films will affect the superconductivity\cite{Nabe18.FeSeStrain} (and may also affect the magnetism), detailed systematic research will be needed to clarify the relation between the magnetism and the superconductivity in the \fss films. 
Another important subject is whether the magnetism coexists or compete with the superconductivity, which also remains for further studies.

The appearance of the magnetism after the suppression of the structural transition in the FeSe$_{1-x}$S$_x$ films is very similar to what was observed in bulk FeSe under pressure\cite{NatCom.7.12146}.
This may suggest that the magnetism observed in the FeSe$_{1-x}$S$_x$ films has the same feature as that for FeSe under pressure.
However, the $T_{\mathrm c}$ behavior is significantly different between these two systems.
The $T_{\mathrm c}$ values of FeSe$_{1-x}$S$_x$ films continues to decrease in the magnetic phase up to $x \sim 0.45$, while $T_{\mathrm c}$ gradually increases with increasing pressure deep in the magnetic phase in FeSe under hydrostatic pressure.
A theoretical study suggests that various stripe and N\'eel states compete with each other and FeSe lies near a multi-critical point in the magnetic phase diagram\cite{Glasbrenner.NaturePhys.11.953}.
In addition, a \musr study reported a magnetic quantum critical behavior in FeSe\cite{PhysRevB.97.201102}.
These studies suggest a possibility that the magnetic order in the S-substituted FeSe films is different from that in FeSe under pressure. %, which results in the different $T_{\mathrm c}$ behaviors in these systems.

We compare the magnetic states in the two systems in terms of the internal magnetic field.
A \musr study with a bulk single crystal suggests that the magnetic state in FeSe under pressure is either the collinear or bi-collinear antiferromagnetic order with Fe spins lying in the $ab$ plane\cite{PhysRevB.95.180504}.
This is different from our observations in the FeSe$_{1-x}$S$_x$ films; the LF measurements suggest that the internal field does not lie in the $ab$ plane, even if effects of the domains with different orientations near the surface of the films are considered.
In addition, the estimated internal field of approximately 100 G for $x=0.4$ is smaller than that for FeSe under pressure ($\sim$ 500 G) with the similar magnetic transition temperature ($\sim$ 50 K).
%Although the weaker internal field for the FeSe$_{1-x}$S$_x$ films might suggest the different magnetic order in FeSe under chemical pressure, it can be explained also by the short-range nature of the magnetic order in the FeSe$_{1-x}$S$_x$ films.
%Indeed, a long-range order was suggested in FeSe under physical pressure by a coherent muon precession\cite{PhysRevLett.104.087003,PhysRevB.85.064517}, different from our samples.
These results suggest that the magnetic order in the FeSe$_{1-x}$S$_x$ films is different from that in FeSe under pressure, which may result in the different $T_{\mathrm c}$ behaviors in these systems.

Matsuura $et\ al$.\cite{NatCom.8.1143} reported that the chalcogen heights from the two-dimensional Fe plane are different between physical and chemical pressure, comparing their results in bulk FeSe$_{1-x}$S$_x$ with results in FeSe under hydrostatic pressure by Millican $et\ al$.\cite{MILLICAN2009707}, which may be the origin for the different magnetic orders in these systems. %and they concluded} that the chalcogen heights determines the \tc behaviors of FeSe under physical and chemical pressure.
However, some other groups reported a different behavior of the chalcogen height in FeSe under hydrostatic pressure other than the results of Millican $et\ al$.
Although Millican $et\ al$. reported that the chalcogen height was almost pressure independent, other groups reported that the chalcogen height also decreased with increasing pressure\cite{Marga09,PhysRevB.81.205119}, similar to the behavior for bulk FeSe$_{1-x}$S$_x$.
This calls for further detailed studies on the structural analysis for FeSe$_{1-x}$S$_x$ and FeSe under pressure, as well as on the magnetic states in these systems.
%Therefore, the further comparative investigation of the structural and physical properties between FeSe$_{1-x}$S$_x$ and FeSe under pressure is essentially important.
%This demonstrates that there are missing parameters, other than the chalcogen height, that determine the $T_{\mathrm c}$ values in iron chalcogenides.
%This calls for further detailed comparative investigation of the structural and physical properties between FeSe$_{1-x}$S$_x$ and FeSe under pressure.

Heavily-electron-doped FeSe, which has electron Fermi surfaces alone, shows high \tc of 40-50 K\cite{PhysRevB.82.180520,SciRep.2.426,JPSJ.82.123705,Shi_2018,PRL.116.077002,ShikamaAPEX2020}.
An interesting issue is the relationship between the high \tc superconductivity in the heavily-electron-doped FeSe and the magnetism in pristine FeSe.
The realization of magnetism under ambient pressure will make it possible to track changes in the magnetism over electron-doping by, for example, electric field effects, which might lead to the elucidation of the mechanism of the high \tc superconductivity in the heavily-electron-doped FeSe.

\section{\label{s:conclusion}Conclusion}
We investigated the magnetic properties of FeSe$_{1-x}$S$_x$ thin films on LaAlO$_3$ substrate by means of $\mu$SR. 
We observed a drastic decrease of initial asymmetry together with a peak structure in the temperature dependence of the relaxation rate at almost the same temperatures, where a kink anomaly was also observed in the temperature dependent resistivity.
%With increasing S content, the anomaly temperature increased and the magnetic fluctuations at the lowest temperature were suppressed.
Our results indicate that the S substitution induces magnetism in FeSe$_{1-x}$S$_x$ thin films. 
Although the behaviors of the magnetic and nematic phases toward chemical pressure by S substitution in FeSe film samples resembles to those toward hydrostatic pressure in bulk FeSe, \tc of thin films monotonically decreases with increasing chemical pressure, which is in contrast to the results for hydrostatic pressure. 
Our findings demonstrate that comparing the effect of chemical pressure between bulk and film samples, as well as the effect of physical pressure on FeSe, is indispensable to understand the interplay of the magnetism, the nematicity and the superconductivity in iron chalcogenides.

\begin{acknowledgments}
Part of this work is based on experiments performed at the Swiss Muon Source S$\mu$S, Paul Scherrer Institute, Villigen, Switzerland. We thank the staff of PSI/S$\mu$S for their help with the $\mu$SR experiments. We also  thank K. Ueno at the University of Tokyo for the X-ray measurements. This research was partially supported by JSPS KAKENHI Grant Numbers JP18H04212, JP19K14651, and JP20H05165.
\end{acknowledgments}

\begin{table}[h]
	\caption{\label{tab:FilmSpec}Specifications of the grown FeSe$_{1-x}$S$_x$ films.}
	\begin{ruledtabular}
		\begin{tabular}{ccccccc}
		& \parbox[c]{4em}{$x$} & \parbox[c]{8em}{thickness (nm)}  & \parbox[c]{4em}{$c$ (\AA)}  & \parbox[c]{4em}{$T_{\mathrm {kink}}$ (K)} \\ \hline
		S30\#1 & 0.3 & 50 & 5.37 & 30  \\
		S30\#2 & 0.3 & 80 & 5.38 & 25 \\
		S40\#1 & 0.4 & 60 & 5.30 & 34 \\
		S40\#2 & 0.4 & 100 & 5.32 & 30 \\
		\end{tabular}
	 \end{ruledtabular}
\end{table}

%& \parbox[c]{6em}{$T_{\mathrm c}^{\mathrm {onset}}$ (K)}  & \parbox[c]{4em}{$T_{\mathrm c}^{\mathrm {zero}}$ (K)}

%\appendix
%\section{Appendixes}

\end{document}